\documentclass[12pt,preprint]{aastex}
\newcommand{\myemail}{lychiang@asiaa.sinica.edu.tw}
\newcommand{\bi}[1]{\mbox{\boldmath $#1$}}
\usepackage{float}
\usepackage{amsmath}
\usepackage{epsfig}
\usepackage{subfigure}
\makeatletter

\usepackage{amsmath}
\usepackage{epsfig}
\newcommand{\beq}{\begin{equation}}
\newcommand{\eeq}{\end{equation}}
\newcommand{\be}{\begin{eqnarray}}
\newcommand{\ee}{\end{eqnarray}}

\def\alm{a_{\l m}}
\def\cl{C_{\l}}
\def\ck{C_k}
\def\cmb{{\rm cmb}}

\def\kt{{k_\theta}}
\def\kp{{k_\varphi}}

\def\glesp{G{\sc lesp }}
\def\healpix{H{\sc ealpix }}

\def\planck{{\rm Planck }}
\def\wmap{{\rm WMAP }}

\def\etal{{et al.}}
\def\l{{\ell}}
\def\lm{{\l m}}

\def\ylm{Y_{\l m}}

\def\cmb{{\rm c}}

\def\cl{{C_\l}}

\def\xk{{X_k}}

\slugcomment{Submitted to Astrophysical Journal}

\shorttitle{Direct Measurement of the \cmb power spectrum}
\shortauthors{Lung-Yih Chiang \& Fei-Fan Chen}

\begin{document}
\title{Direct measurement of the angular power spectrum of cosmic microwave background temperature anisotropies in the WMAP Data}
\author{Lung-Yih Chiang}
\affil{\asiaa}
\author{Fei-Fan Chen}
\affil{\ntu}
\affil{\asiaa}

\email{\myemail}
\newcommand{\ntu}{{Institute of Astrophysics, National Taiwan University, 1, Rooservolt Road, Taipei, Taiwan}}
\newcommand{\asiaa}{{Institute of Astronomy and Astrophysics, Academia Sinica, P.O. Box 23-141, Taipei 10617, Taiwan}}

\begin{abstract}
Angular power spectrum of the cosmic microwave background (CMB) temperature anisotropies is one of the most important  on characteristics of the Universe such as its geometry and total density. Using flat sky approximation and Fourier analysis, we estimate the angular power spectrum from an ensemble of least foreground-contaminated square patches from \wmap W and V frequency band map. This method circumvents the issue of foreground cleaning and that of breaking orthogonality in spherical harmonic analysis due to masking out the bright Galactic plane region, thereby rendering a direct measurement of the angular power spectrum. We test and confirm Gaussian statistical characteristic of the selected patches, from which the first and second acoustic peak of the power spectrum are reproduced, and the third peak is clearly visible albeit with some noise residual at the tail. 
\end{abstract}

\section{Introduction}
The angular power spectrum of the cosmic microwave background (CMB) temperature anisotropies contains a wealth of information about the properties of our Universe. The physics behind the shape of the power spectrum at different angular scales is well understood (e.g. see Hu et al.) and it therefore allows us to distinguish different cosmological models. The power spectrum possesses specific features, known as acoustic peaks, characterizing compression and rarefaction of the photon-baryon fluid around the decoupling epoch. NASA Wilkinson Microwave Anisotropy Probe (\wmap) \citep{wmap1yrresult,wmap3yrcos,wmap5yrresult,wmap7yrresult} has produced results that has ushered in the era of ``Precision Cosmology'', including the angular power spectrum \citep{wmap1yrpower,wmap3yrtem,wmap5yrpower,wmap7yrpower}, from which cosmological parameters are estimated to a high precision \citep{wmap1yrcos,wmap3yrcos,wmap5yrcos,wmap7yrcos}.

In order to retrieve the angular power spectrum, however, one has to separate the foreground contamination from our own Galaxy and extra-galactic point sources in the observed data \citep{wmap1yrfg,wmap3yrtem,wmap5yrfg,wmap7yrfg}. The standard treatment of eliminating Galactic diffuse foreground is via multifrequency cleaning, via minimum variance optimization for extracting the angular power spectrum, or through known foreground templates. Another issue arising from the foreground contamination is the strong emission of the Galactic plane, for which various masks are adopted by \wmap science team. Masking procedure and incomplete sky coverage thus breaks the orthogonality in the spherical harmonic analysis, which requires additional attention in obtaining the spherical harmonic coefficients \citep{master,oh,maskbias}. 

Apart from \wmap science team, only a couple of papers are devoted to extraction of the CMB power spectrum from raw data \citep{saha05,saha07,samal,nilc}. They all adopt the methodology of internal linear combination and implement quadratic minimization, which not only minimizes the foreground contamination, but also subtracts the power that is related with the chance correlation between the CMB and foregrounds \citep{covariance}.

In this paper we present a simple method of direct measurement of the CMB angular power spectrum from \wmap raw data, the frequency band maps. By ``direct'' we mean circumventing any foreground subtraction techniques and avoiding the issue of incomplete sky coverage. We also use \wmap frequency band maps, which are made possible  for power spectrum extraction after \citet{xps} estimate the corresponding window functions.

This paper is arranged as follows. In Section 2 we review the flat sky approximation and then we discuss the issue of foregrounds, instrument noise and window function in Section 3. We test the Gaussianity of the patches taken from WMAP data in Section 4. We then employ the method in Section \ref{realdata}, and the Discussion is in Section \ref{discussion}.

\section{Flat sky approximation}

\begin{figure}
\plotone{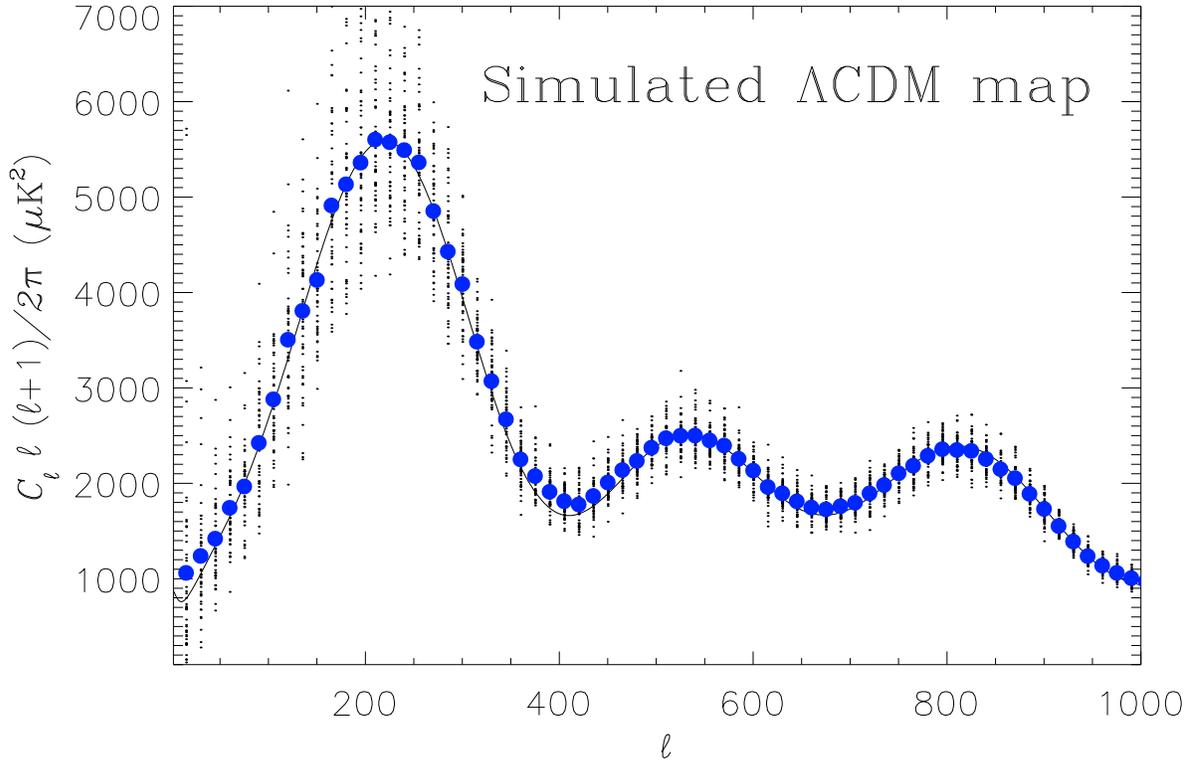}
\caption{Test of the scaling relation Eq.(\ref{relat}) in flat sky approximation. We simulate a full sky CMB map with WMAP best-fit $\Lambda$CDM model and take 50 patches of $24^\circ \times 24^\circ$ square with pixel size 3 arcmin. The power spectra of the patches are scaled with a factor $(2\pi/15)^2$ and the sampling interval of the  multipole numbers $\Delta \l=15$. The dots are from 100 patches and the mean is denoted by big blue dots. One can see the mean power spectrum after scaling fits nicely with the input one and the issue of discontinuous boundary condition is negligible.}
\label{scaling}
\end{figure}

Standard treatment for whole-sky CMB spectral analysis is via writing the temperature anisotropies as a sum of spherical harmonics $\ylm$: $T(\theta,\varphi)=\sum_\l \sum_m \alm \ylm(\theta,\varphi)$, where $\theta$, $\varphi$ are the polar and azimuthal angle, $\alm\equiv |\alm| \exp(i\phi_\lm)$ is the spherical harmonic coefficient and $\phi_\lm$ is the phase.  The strict definition of an isotropic GRF requires the real and imaginary part of the $\alm$ mutually independent and both Gaussian, but a more convenient definition is that the phases are uniformly random on the interval $[0,2\pi]$. The power spectrum can be estimated $\cl=(2\l+1)^{-1}\sum_m |\alm|^2$. 

To estimate the power spectrum from small square patches, however, one can use Fast Fourier Transform (FFT): 
\begin{equation}
T({\bi r})=\sum_{\bi k} a_{\bi k} \exp\left[\frac{2\pi i( {\bi r} \cdot {\bi k})}{N}\right],
\end{equation}
where ${\bi r}\equiv(\theta, \varphi)$ and ${\bi k}\equiv (\kt, \kp)$ if the patches are chosen on the equator with sides aligned with the spherical coordinates. The power spectrum from the patch is $C_k\equiv \langle|a_{\bi k}|^2 \rangle$, where the angle brackets denote average for integer $k$ over all $ |a_{\bi k}|^2$ for $k-1/2 \le |{\bi k}| < k+1/2$. The scaling relation between Fourier wavenumber $k$ and multipole number $\l$ is $\l=2 \pi k /L$, where $L$ is the patch size. The angular power spectrum $\cl$ at multipole number $\l$ is scaled from $C_k$ at Fourier wavenumber $k$ via
\begin{equation}
C_{\l= 2\pi k/L}=  L^2 C_k.
\label{relat}
\end{equation} 
The scaling relation can be easily understood as follows if the signal is white noise. For spherical harmonic analysis $\cl= 4 \pi \sigma^2_{\rm sky}/ N_{\rm sky}$ whereas for FFT on a patch taken from the sky $\sigma^2_{\rm patch} =  C_k N_{\rm patch}$, where $N_{\rm sky}$ is the total pixel number of the sphere and $N_{\rm patch}$ pixel number of the patch. Since white noise is homogeneous, $\sigma^2_{\rm patch}=\sigma^2_{\rm sky}$, then $\cl=4 \pi C_k N_{\rm patch}/N_{\rm sky}= C_k L^2$. Note that the scaling relation can be applied with minimum error for patches centred at $\theta=\pi/2$ if one uses non-equal area pixelization scheme. 

According to the scaling relation, the largest scale (smallest $\l$) at which one can obtain the power is $\l_{\min}= 2 \pi/L$\footnote{Usually one associates multipole number $\l$ to a characteristic angular scale $\varpi$ on the sphere via $\l=\pi/\varpi$ because a characteristic angular scale (e.g. a scale between a cold and a hot spot) is half of one full wavelength, i.e. $\varpi=L/2$.}, then the multiple numbers are sampled with the interval $\Delta \l= 2 \pi/L$ down to the smallest scale decided by the size of the pixel $p$ : $\l_{\rm max}= \pi/p$. So the disadvantage of estimation of power spectrum from patches is that sampling interval $\Delta \l$ is much larger than 1, which can be viewed as intrinsic binning.

In Figure \ref{scaling} we test the scaling relation of Eq.(\ref{relat}). We simulate a full sky CMB map with WMAP best-fit $\Lambda$CDM model and take 50 patches, each $24^\circ \times 24^\circ$ with pixel size 3 arcmin.  One can see that the mean power spectrum from the 50 patches fits nicely with the input power spectrum and the error from the discontinuous boundary condition usually present in data analysis of square patches is negligible.

\section{Directly Retrieving the CMB power spectrum}
The signal $T_\nu$ in the sky at frequency $\nu$ is a combination of the CMB signal $T_\cmb$ and diffuse foregrounds (synchrotron, free-free and dust emission) plus extragalactic point sources, altogether denoted as total foreground $F_\nu$. They are measured with an antenna beam $B_\nu$: 
\begin{equation}
T_\nu=(T_\cmb + F_\nu) \otimes B_\nu + N_\nu,
\label{general}
\end{equation}
where $\otimes$ denotes convolution and $N_\nu$ is the instrument noise. In order to reach the CMB power spectrum, we discuss below the 3 parts in Eq.(\ref{general}): foreground contamination, noise and the window function.
 
\begin{figure}
\plotone{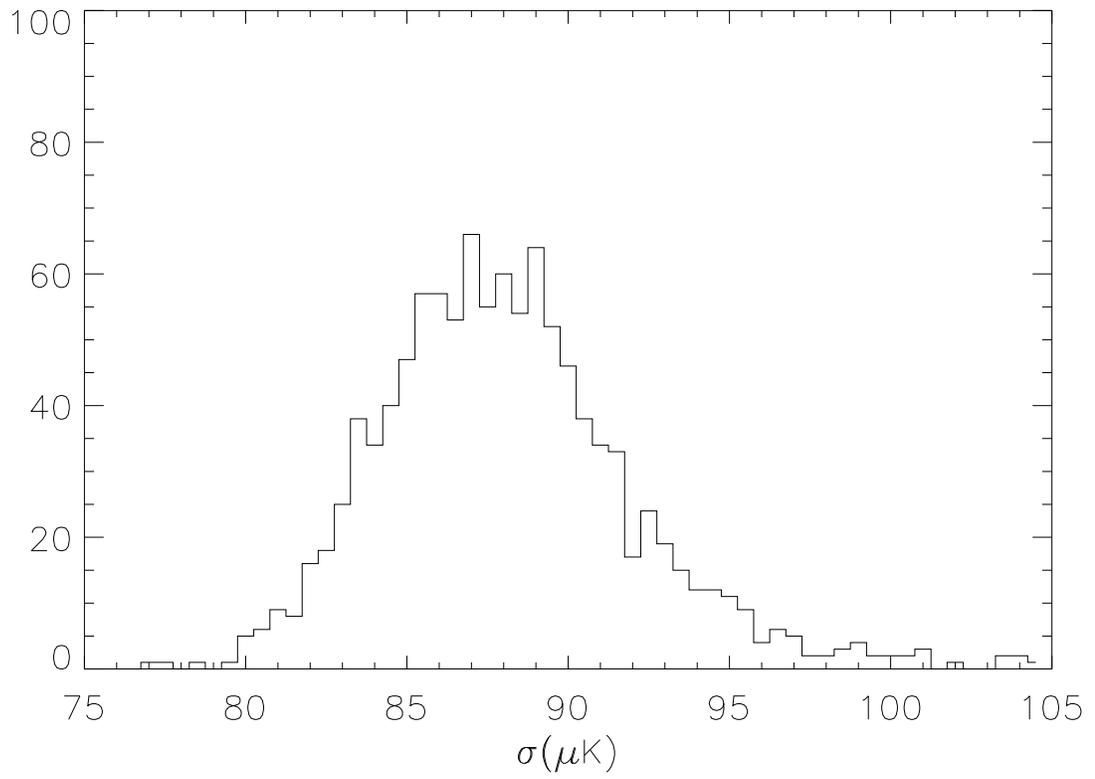}
\caption{The histogram of CMB fluctuation from 1000 $24^\circ \times 24^\circ$ patches, which are taken from simulated full-sky maps with beam FWHM 19 arcmin. The mean is at $88.34\mu$K.}
\label{sim}
\end{figure}

\subsection{Foreground contamination and dispersion threshold} 
NASA Cosmic Background Explorer has measured with $10^\circ$ FWHM the CMB temperature fluctuation at a level $10^{-5}$ \citep{coberms}.  From Eq.(\ref{general}) the variance of the measured $T_\nu$ includes foreground component: $\sigma^2_\nu= \sigma^2_\cmb + \sigma^2_{F_\nu}+ \sigma^2_n+{\rm Cov}[T^{\rm sm}_\cmb,F^{\rm sm}_\nu]+{\rm Cov}[T^{\rm sm}_\cmb,N_\nu]+{\rm Cov}[F^{\rm sm}_\nu,N_\nu]$, where $\sigma_\cmb^2$, $\sigma_{F_\nu}^2$ and $\sigma_n^2$ are the variance of the beam-convolved CMB, beam-convolved foreground and noise at frequency $\nu$, respectively, and the last 3 terms denote their covariances. For an ensemble of small patches, the average  
\begin{equation}
\langle \sigma^2_\nu \rangle = \langle \sigma^2_\cmb \rangle + \langle \sigma^2_n \rangle  + \langle \sigma^2_{F_\nu} \rangle  \ge \langle \sigma^2_\cmb \rangle + \langle \sigma^2_n \rangle.
\label{crit}
\end{equation}
The CMB fluctuations (and noise) always persist in each patch, but those from the foreground do not. Thus we can choose patches with lower variances, as they contain less foreground contamination, thereby providing better estimation of the CMB power spectrum. One can therefore use dispersion threshold $\sigma^{\rm th}$ for controlling foreground contamination level among patches.

\begin{figure}
\epsfig{file=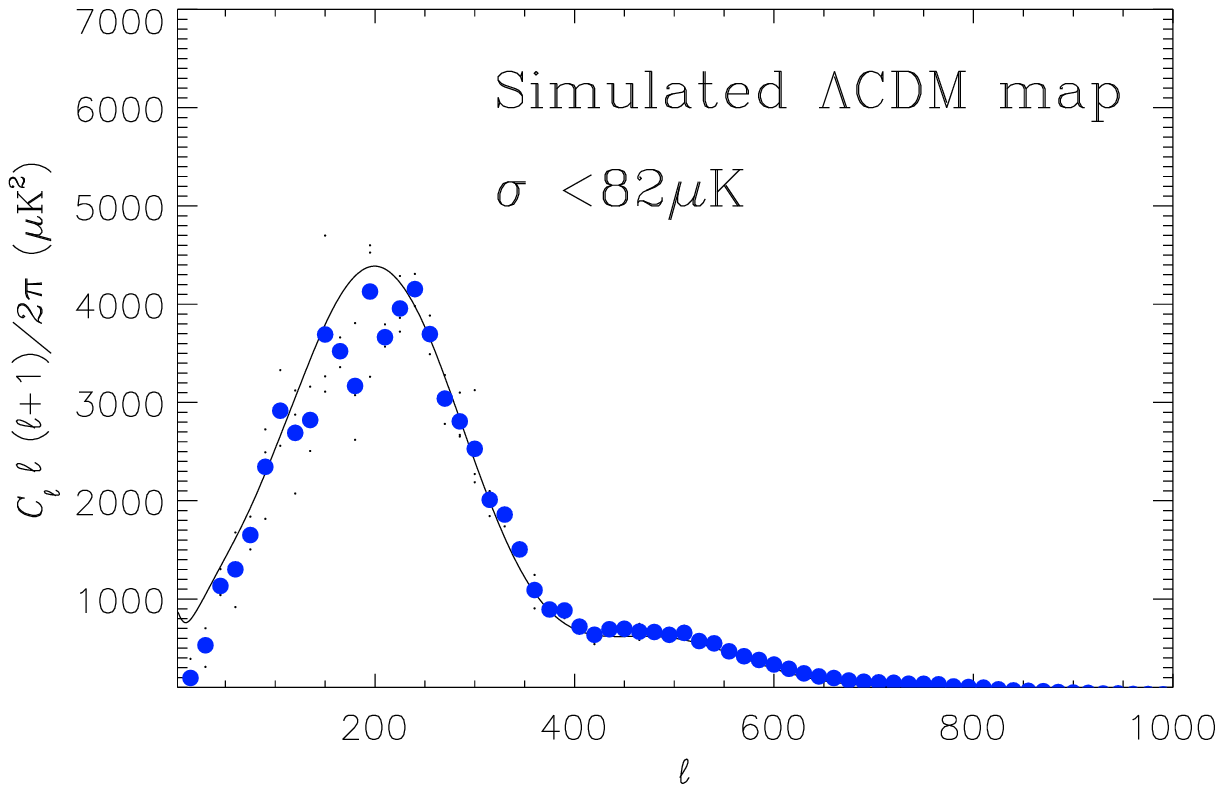,width=5cm}
\epsfig{file=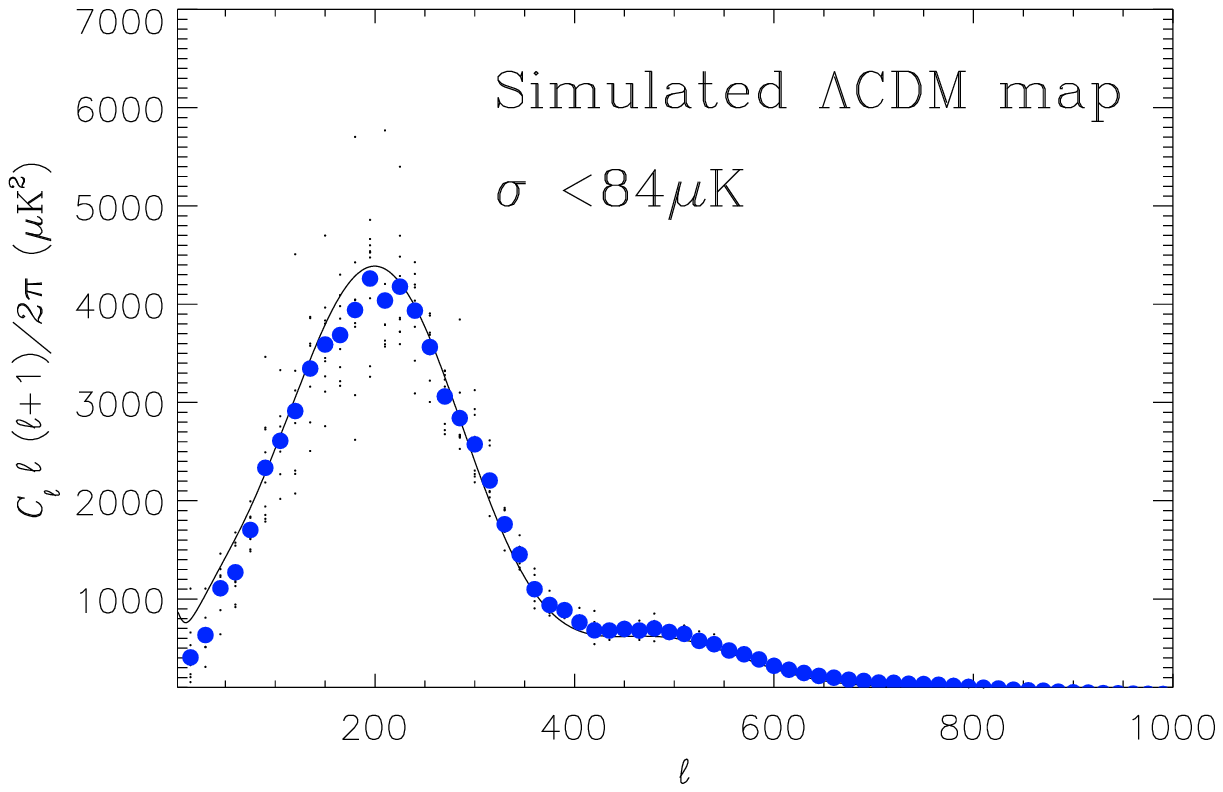,width=5cm}
\epsfig{file=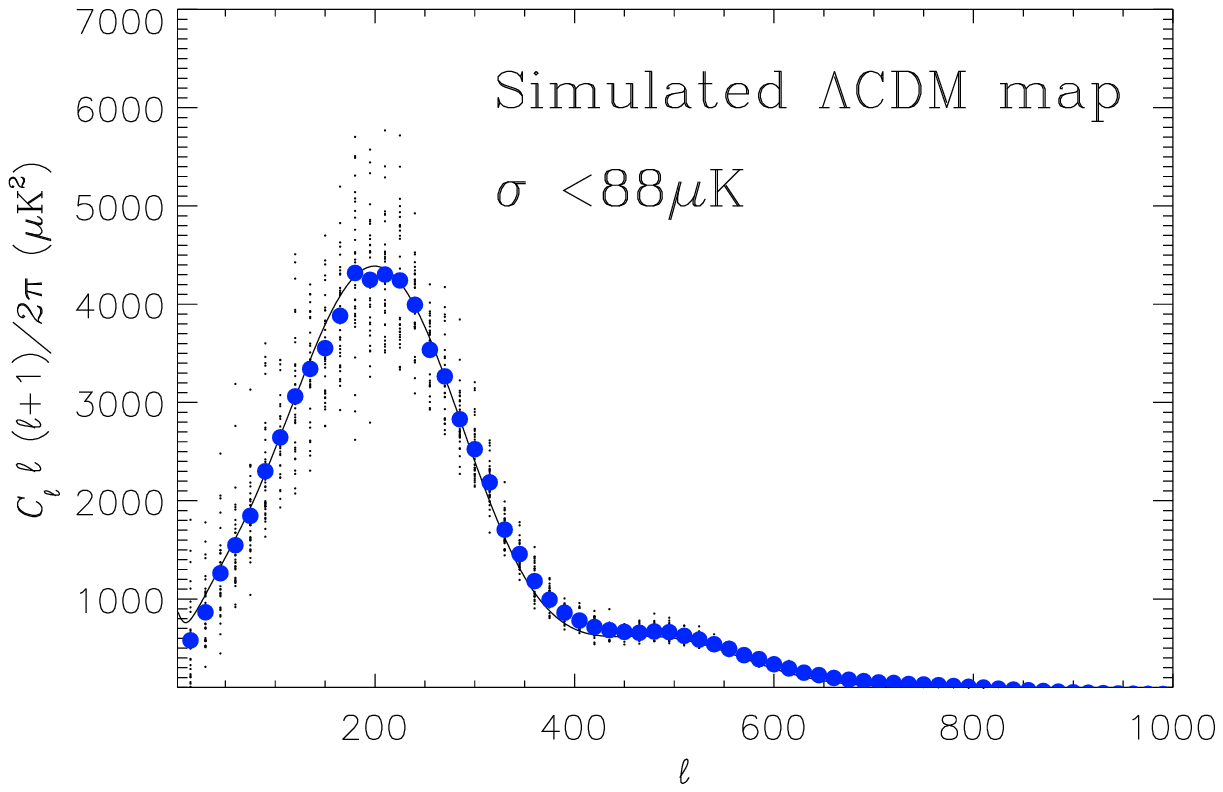,width=5cm}
\caption{Demonstration of CMB quiet areas from low variance. From a simulated V band CMB map (high $\l$ are smoothed by beam FWHM 21 arcmin), we set different threshold $\sigma^{\rm th}$ to see if choosing CMB quiet areas affects the power spectrum estimation. On the left panel we choose threshold $\sigma^{\rm th}=82\mu$K with only 3 patches in the plot. The mean power spectrum (big blue dot) is indeed lower than the input one (solid line), particularly for low $\l$. In the middle panel we plot the 9 patches that have $\sigma$ lower than 84$\mu$K. On the right panel with $\sigma^{\rm th}=88\mu$K (the mean from Fig.\ref{sim}), one can see the mean power spectrum from 34 patches fits well with the input one.}
\label{quiet}
\end{figure}

Power spectrum is a spread of the variance into different scales, so from Eq.(\ref{crit}) $\langle \ck^\nu \rangle=(\langle\ck^\cmb \rangle+\langle f_k^\nu \rangle)\mathcal{W}_k^\nu +\langle n_k^\nu\rangle$, where $\ck^\cmb$ is the power spectrum of the CMB,  $\mathcal{W}_k^\nu$ is the window function, $\ck^\nu$, $f_k^\nu$ and $n_k^\nu$ are the power spectrum of  the band signal, total foreground and instrument noise of frequency $\nu$ at wavenumber $k=\l L/2 \pi$,  respectively. For \wmap V and W band where the CMB dominates over the foreground outside the Galactic plane, patches with variance below a threshold $\sigma_\nu^2 < \sigma_{\rm th}^2$ shall give
\begin{equation}
\langle \ck^\nu \rangle \simeq \langle \ck^\cmb \rangle \mathcal{W}_k^\nu + \langle n_k^\nu\rangle.
\label{cl}
\end{equation}

We simulate WMAP V band (i.e. with beam FWHM 21 arcmin) full-sky CMB maps and take in total 1000 patches of $24^\circ \times 24^\circ$ square and plot in Fig.\ref{sim} the histogram of the dispersion $\sigma$. The mean lies at $88.34 \mu$K, which provides an indication of our choice of the dispersion threshold.

There is concern that by choosing low-variance patches we are choosing CMB quiet area, which might result in lower power spectrum. In Fig.\ref{quiet} we demonstrate that unless a significantly low threshold is chosen (which would result in few patches), using low variance as a criterion still provides a fair sample for power spectrum estimation.

\subsection{Cross-power spectrum to eliminate noise}
To eliminate the noise after choosing patches with low variance, we can employ cross-power spectrum on the same patch of the sky at different frequency bands. Cross-power spectrum (XPS) is a quadratic estimator between two maps (or patches) $a$ and $b$, whose Fourier modes are $a_{\bi k}$ and $b_{\bi k}$:
\begin{equation}
x_k^{ab}= \frac{1}{2} \langle ( a^{*}_{\bi k} b_{\bi k} + b^{*}_{\bi k} a_{\bi k})\rangle,
\label{def}
\end{equation}
where $*$ denotes complex conjugate and the angle brackets have the same notation as in Eq.(\ref{relat}).  The advantage of XPS as an unbiased quadratic estimator for power spectrum estimation lies in the fact that XPS returns with its usual power spectrum $\langle |a_{\bi k}|^2 \rangle$ if $a$ and $b$ are of the same signal. If $a$ and $b$ are uncorrelated then XPS reduces the signal by \citep{xps}
\begin{equation}
\frac{\sqrt{\langle (\xk^{ab})^2 \rangle} }{\sqrt{ A_k B_k }}\simeq \frac{1}{\sqrt{2\pi k}},
\label{xpsuncorr}
\end{equation}
where $A_k$ and $B_k$ is the power spectrum of signal $a$ and $b$ respectively. The decreasing of the uncorrelated signal is inversely proportional to the square root of the number of random walk, and it can be further decreased by $1/\sqrt{LN}$ with binning $L\equiv \Delta \l$ multipole numbers and averaging from $N$ sets of XPS. Therefore XPS is useful in reducing uncorrelated signals while preserving the correlated one, which is employed by WMAP to extract CMB spectrum by crossing the foreground-cleaned maps from Differencing Assemblies (DA) \citep{wmap1yrpower,wmap3yrtem,wmap5yrpower,wmap7yrpower}.

For patches on V and W band map with low variances, hence satisfied Eq.(\ref{cl}), we can write $a_{\bi k}^{\rm V}=a_{\bi k}^{\rm c}b_{\bi k}^{\rm v}+ n^{\rm v}_{\bi k}$ and $a_{\bi k}^{\rm W}=a_{\bi k}^{\rm c}b_{\bi k}^{\rm w}+ n^{\rm w}_{\bi k}$, where $a_{\bi k}^{\rm c}$ is the Fourier mode of CMB, $b_{\bi k}^{\rm v}$ and $ b_{\bi k}^{\rm w}$ are that of V and W band beam, and $n^{\rm v}_{\bi k}$ and $n^{\rm w}_{\bi k}$ that of V and W band noise, respectively. In XPS the correlated signal $\langle|a^{\rm c}_{\bi k}|^2 b_{\bi k}^{\rm v} b_{\bi k}^{\rm w} \rangle$ is what we look for whereas those uncorrelated terms between CMB and noises $\xk^{\rm cw}$, $\xk^{\rm cv}$ and between noises $\xk^{\rm vw}$ shall be decreased according to Eq.(\ref{xpsuncorr}).

\subsection{Window Functions of the Frequency Band Maps}
The window functions of the WMAP DA maps are directly measured from Jupiter \citep{wmap1yrbeam,wmap5yrbeam} and are available at the official website \footnote{http://www.lambda.gsfc.nasa.gov/product/map/current/}. The frequency band maps, however, are combined from the DA maps, so the corresponding window functions do not exist. Note that the window functions of the DA maps even at the same frequency band have different profiles, particularly for the W frequency band. It is then demonstrated in \citet{xps} that the window functions of the frequency band maps can be estimated from bright point sources and it is shown that the window function of the W band map takes the form of that of W1 DA, whereas that of V band takes that of V1 or V2 DA. 

\section{Gaussianity of the patches}
The simplest inflation theory predicts the CMB anisotropies, amplified from quantum fluctuations, constitute a Gaussian random field (GRF) \citep{bbks,be}. If the CMB is indeed statistically isotropic Gaussian, the angular power spectrum furnishes a complete statistical description. One of the properties of GRF is its phases from harmonic analysis are uniformly random in $[0, 2\pi]$. Based on the random phase hypothesis we can test Gaussianity of the selected patches by employing the Shannon entropy of Fourier phases ${\cal S}=-\sum p_i \ln p_i \delta \phi$, where $ p_i \delta \phi$ is the distribution probability at the $i$th interval in $[0, 2\pi]$ and $\sum p_i \delta \phi=1$. It can be used to test for uniformity: $p\equiv p(\phi_k)$ and independence (non-association): $p\equiv p(D)$ where $D(\Delta k)=\phi_{k+\Delta k}-\phi_k$ \citep{phaseentropy,nature}. If the phases are uniformly random, then $p_i=(2\pi)^{-1}$ and the entropy reaches a maximum ${\cal S}_{\rm max}=\ln 2 \pi$. 

\begin{figure}
\plottwo{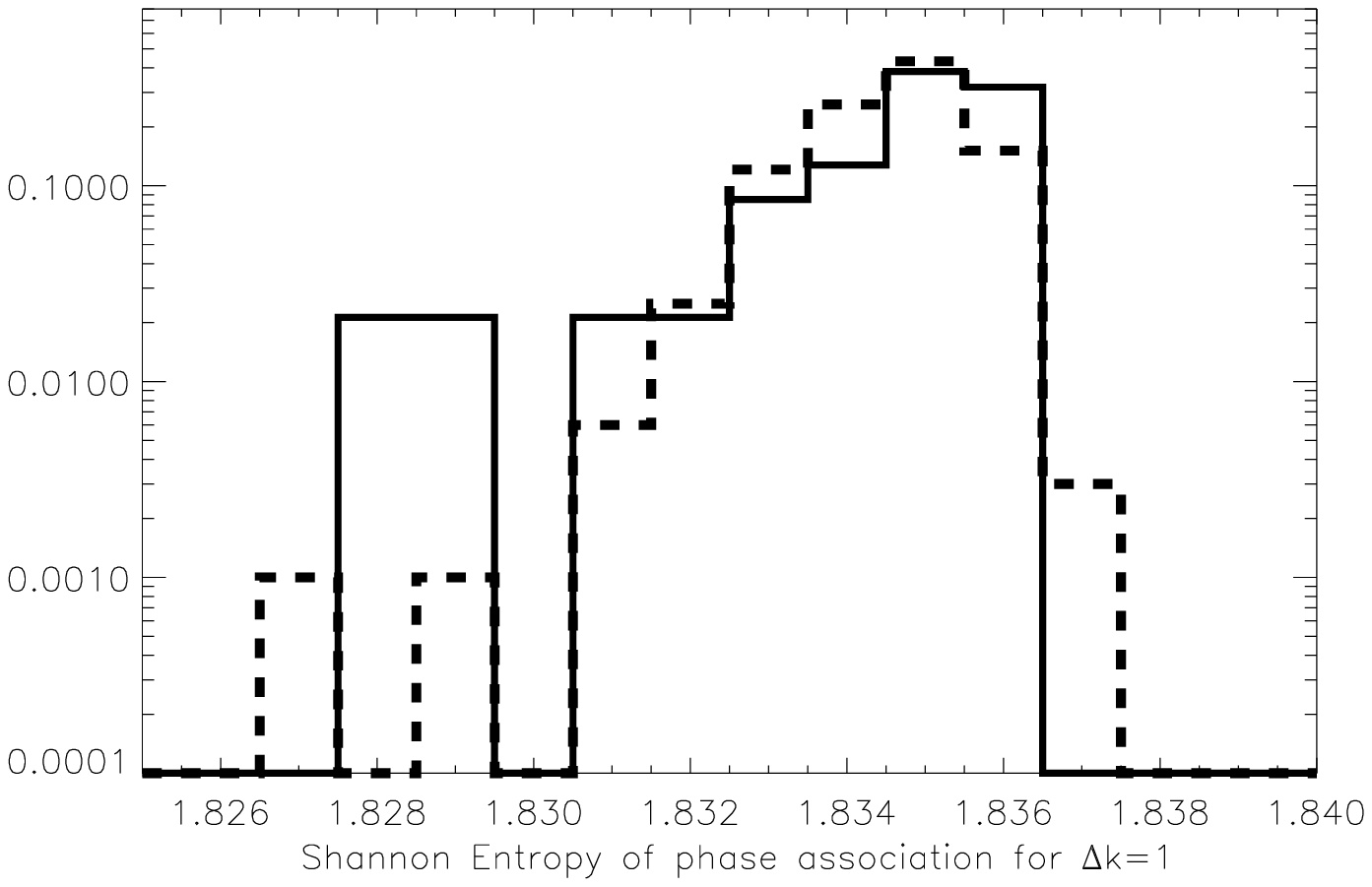}{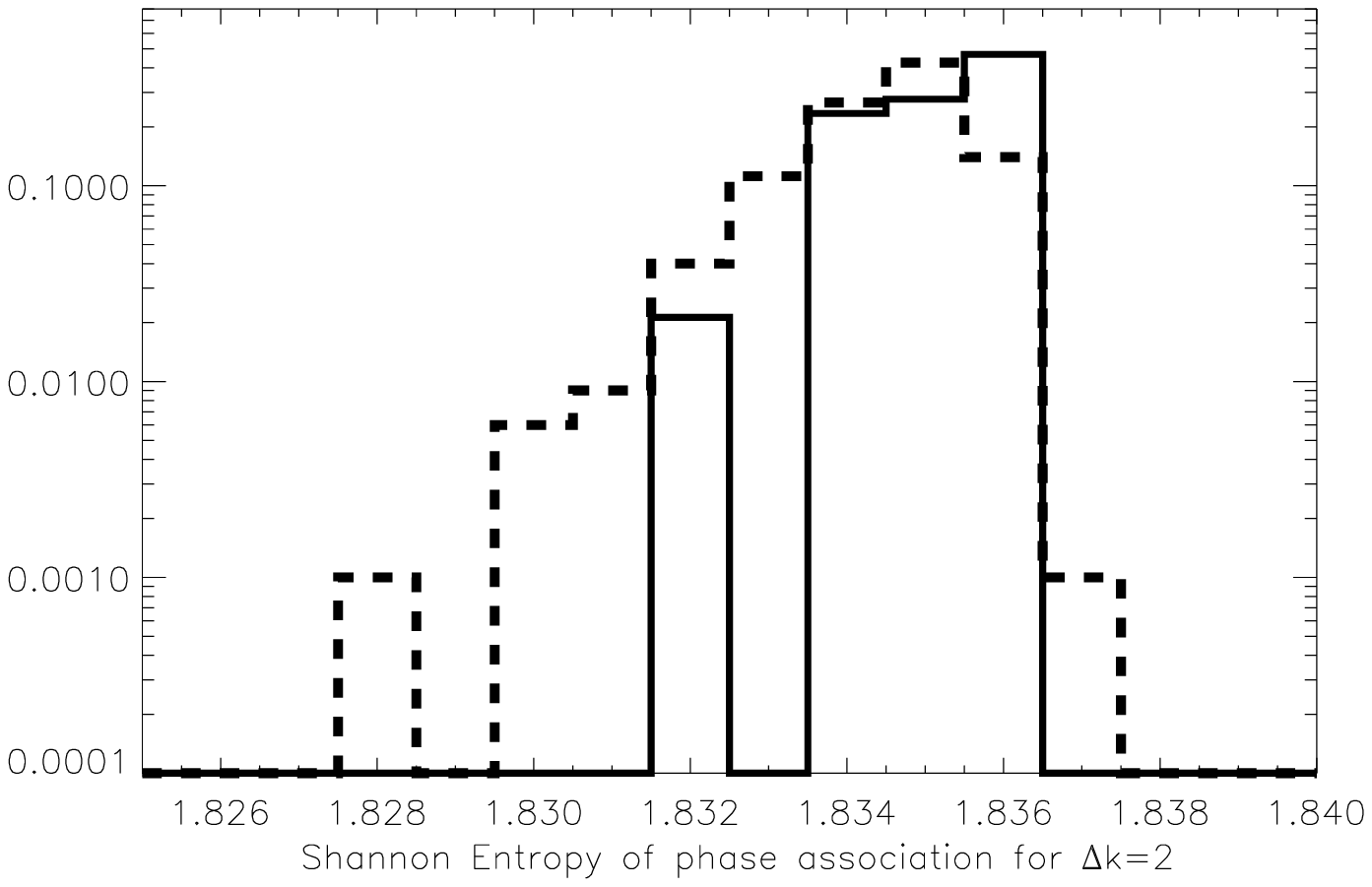}
\caption{Normalized histogram of the Shannon entropy for Fourier phase association from the 47 selected patches. The phases of the Fourier modes equivalent to multipole number $\l \le 1050$ are taken for calculation. The Shannon entropy ${\cal S}$ of phase association between $\Delta k=1$  is shown on the left panel and $\Delta k=2$ on the right. the normalized histogram for the 47 patches is shown in solid curves. For comparison we plot in dash curve the histogram from 1000 $24^\circ \times 24^\circ$ patches taken from full-sky Gaussian maps. }\label{phase}
\end{figure}
\section{Angular power spectrum of the CMB from the WMAP frequency band maps} \label{realdata}
\begin{figure}
\plotone{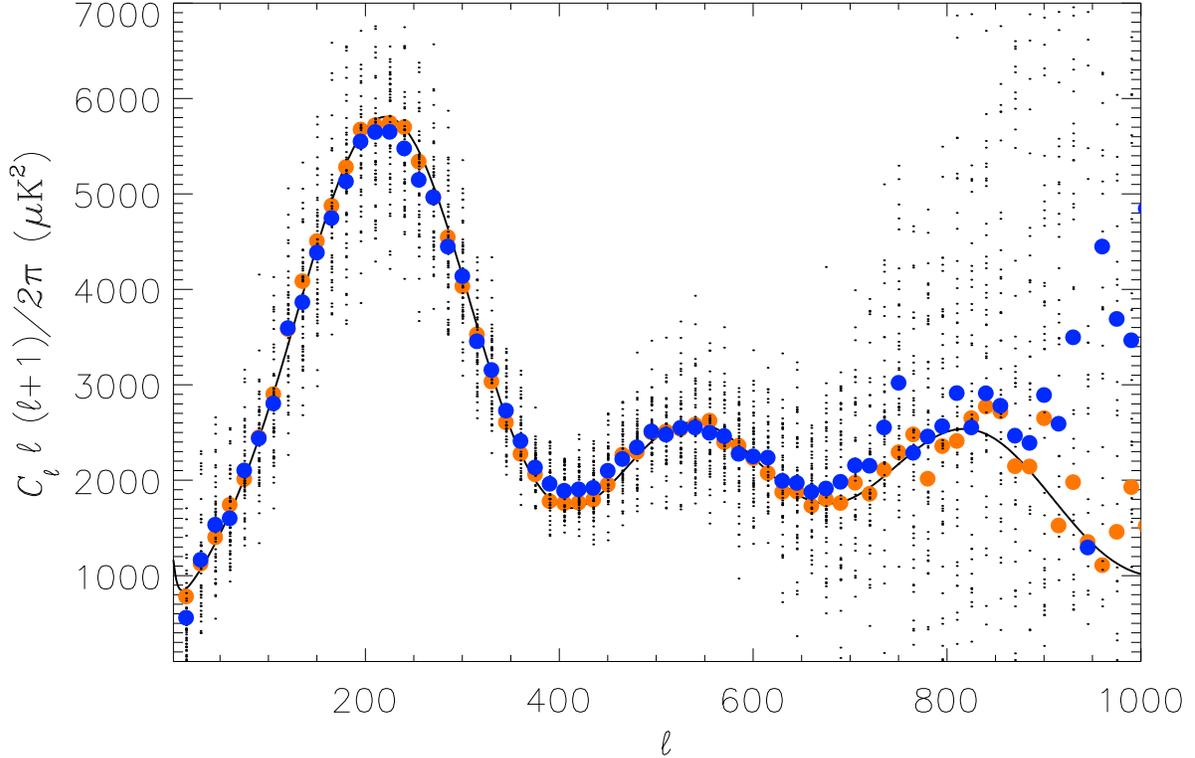}
\caption{Direct measurement of the CMB angular power spectrum. From WMAP V band map we choose patches with $\sigma < 98 \mu $K (after eliminating bright point sources), and we take the cross-power spectra  of patches between WMAP V and W band. After deconvolution of the window functions, the power spectra of the 47 patches are shown in black dot and the mean power spectrum is in big blue dot. For comparison we plot in big orange dot the power spectrum binned ($\Delta \l=15$) from that by the WMAP science team. The best-fit $\Lambda$CDM model is in solid line.}
\label{result}
\end{figure}

In this section we apply our method on \wmap frequency V and W band maps to extract the CMB angular power spectrum via Fourier analysis on $24^\circ \times 24^\circ$ patches. Although one can take patches with a smaller size rendering more patches from the whole sky, it would increase the noise power spectrum level, and consequently, the XPS residual shown in Eq.(\ref{result}).

We first take WMAP V band and choose 47 patches with $\sigma < 98 \mu $K (after deleting bright point sources exceeding 5$\sigma$ of the patch). Note that in Fig.\ref{sim} the mean of the 1000 patches is $\sigma= 88.34\mu$K, but those taken in real maps with pixel noise have higher values. Before extracting the power spectrum, we test the Gaussianity of the 47 patches by employing Shannon entropy of Fourier phases for their association. In Fig.\ref{phase} we show the normalized histogram of Shannon entropy for association of phases for $\Delta k=1$ on the left and 2 on the right. We also plot the histogram from 1000 $24^\circ \times 24^\circ$ Gaussian patches.

We then calculate the XPS from the same patches of the V and W band to eliminate the noise. They are  then deconvolved by $(\sqrt{W_{\rm V} W_{\rm W}})^{-1}$, where $W_{\rm V}$ and $W_{\rm W}$ are the window function of the V and W frequency band map, respectively. We show in Fig.\ref{result} the retrieved angular power spectrum. One can see our simple method yields the CMB power spectrum with clear 1st and 2nd Doppler peak, which match the result from WMAP science team. The 3rd peak is also visible albeit with higher amplitude at the tail than theirs, which is due to the residual from XPS.

\section{Discussion} \label{discussion}
Since WMAP data release, full-sky analysis has become the standard way for power spectrum estimation as it has offered for the first time detailed measurement at super-horizon scales. The method we present in this paper utilizes small patches, hence has intrinsic limitation on the largest scale we can measure.  It nevertheless adopts a totally different methodology and provides a more intuitive way to obtain the power spectrum on all but the very large scales. This method can be readily applied on the upcoming \planck data.

\acknowledgments
We acknowledge the use of \healpix\footnote{{\tt http://healpix.jpl.nasa.gov/}} package \citep{healpix}  and the use of \glesp\footnote{{\tt http://www.glesp.nbi.dk/}} package. The author would like to thank Peter Coles, Andrew Jaffe, Dipak Munshi and Keiichi Umetsu for useful discussions and suggestions.

\expandafter\ifx\csname natexlab\endcsname\relax\def\natexlab#1{#1}\fi
\newcommand{\combib}[3]{\bibitem[{#1}({#2})]{#3}} 

%
%
\newcommand{\autetal}[2]{{#1,\ #2., et al.,}}
\newcommand{\aut}[2]{{#1,\ #2.,}}
\newcommand{\saut}[2]{{#1,\ #2.,}}
\newcommand{\laut}[2]{{#1,\ #2.,}}

%
%
\newcommand{\refs}[6]{#5, #2, #3, #4} 
\newcommand{\unrefs}[6]{#5 #2 #3 #4 (#6)}  

%
%

\newcommand{\book}[6]{#5, #1, #2, #3}
%

\def\apj{ApJ}
\def\apjl{ApJL}
\def\mn{MNRAS}
\def\nature{Nature}
\def\aa{A\&A}
\def\prl{Phys.\ Rev.\ Lett.}
\def\prd{Phys.\ Rev.\ D}
\def\pr{Phys.\ Rep.}
\def\ijmpd{Int. J. Mod. Phys. D}
\def\jcap{J. Cosmo. Astropar.}

\end{document}